# How the Moon Impacts Subsea Communication Cables

Lothar Moeller

SubCom, Eatontown, NJ 07724, USA, lmoeller@subcom.com

**Abstract:** We report tidal-induced latency variations on a transpacific subsea cable. Week-long recordings with a precision phase meter suggest length changes in the sub-meter range caused by the Poisson effect. The described method adds to the toolbox for the new field 'optical oceanic seismology'.

## 1. INTRODUCTION

A novel method based on utilizing the architecture of subsea cables for oceanic seismology has recently been explored experimentally[1],[2]. It is conceptually distinguishable from previous techniques for remote sensing in which fiber serves as the transport medium between underwater sensor arrays[3,4] or is analysed using distributed acoustic sensing (DAS) to monitor the cable performance[5].

In its most mature form, the method can sense large geographic areas since it uses the entire cable length as an aperture and is compatible with simultaneous operation of commercial traffic on the same fiber[2]. This feature makes it universally deployable on existing routes for low-cost and long-term surveillance purposes.

The two implementations of this method, which were tested on subsea earthquakes localization, share the same basic principle of detecting optical phase distortions caused by mechanical stresses and strain on the cable. One version interferometrically resolves phase distortions induced by cable motion and tension[1]. The second version analyses the state of polarization (SOP) fluctuations of data channels logged by modern coherent transponders[2]. While the exact opto-mechanical coupling between a vibrating sea floor and cabled fiber is still being investigated, it is known empirically that the SOP of light in SSMF is sensitive to birefringence changes caused by fiber micro-bending or motion.

Here we report, for the first time, on cable length variations in the sub-meter range that depend on water pressure generated by tidal variations. In contrast to the two aforementioned implementations, the cable is not impacted by any kind of abrupt sea floor motions. However, changes in local water pressure cause cable length changes that are detectable with an ultra-stable phase meter. In contrast to commonly held views, subsea cables diverge from the 'loose tube' model that suggests a forceless bearing on the fibers. Our observations indicate a strong coupling between the cable jacket and the sheathed fiber.

## 2. GPS LONG-TERM STABILIZED RF PHASE METER

The key idea behind our experiment is to assess path length changes by measuring the phase delay of an RF signal that propagates on an optical carrier through the cable. An ultra-stable phase meter compares the arrival phase of an RF signal at the cable output with a stable local reference (btb) as sketched in Fig.1a.

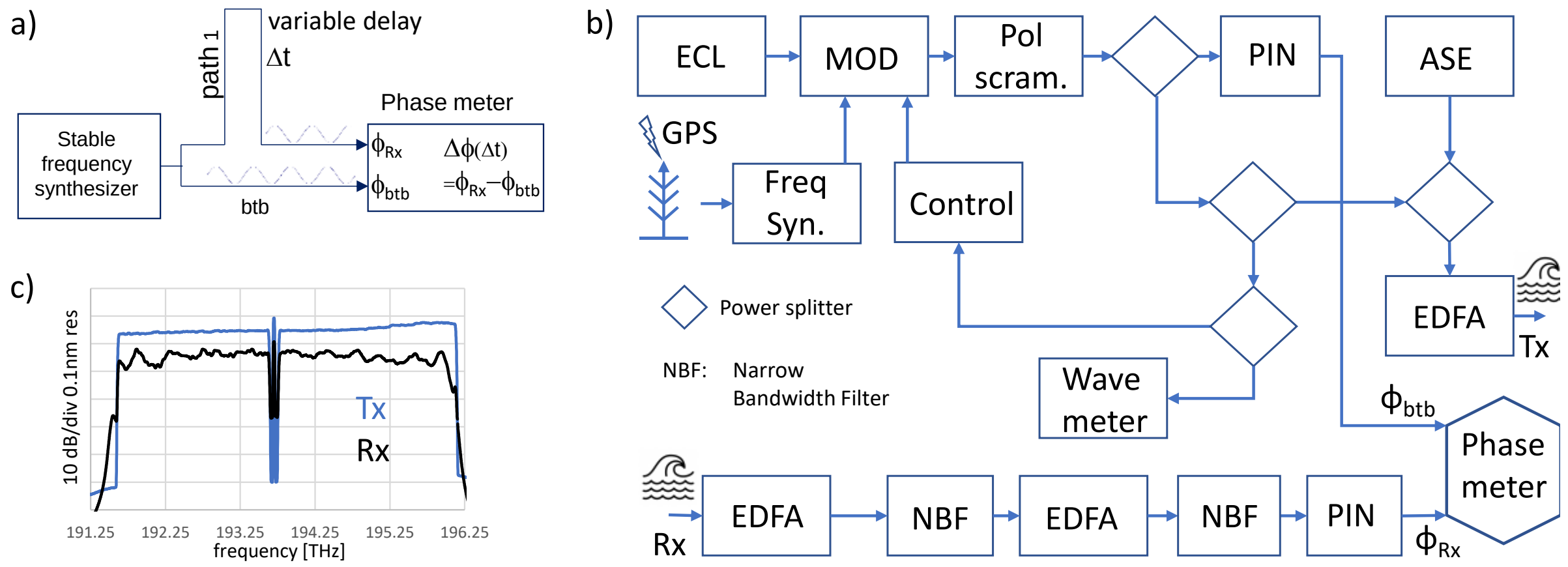

Fig.1: a) Concept of phase meter, b) Tx/Rx setup at Japan. An AM probe @ 20 MHz embedded in ASE loading enters wet plant. Received probe is filtered, amplified, and O/E converted. Phase meter records RF phases of probe at input and output of wet plant. c) Launch and receive spectra.

Any delay variation in path 1 would be captured as a varying phase difference. Contrariwise, a varying phase difference does not automatically point to a length fluctuation; but could also originate from a frequency drift in the RF signal. Typically, signal propagations last over ~100 ms and recordings run weeks. Therefore, a correspondingly stable frequency synthesizer is essential in these experiments. Our Rubidium-controlled OCXO synthesizer locks to a 10-MHz standard, derived from a GPS signal and ensures excellent short-, medium-, and long-term stability. The synthesizer drives an optical $LiNbO_3$ (Fig.1b) modulator connected via a PM fiber to a cw external cavity laser (ECL). A weak optical feedback keeps it locked to the null of its characteristic curve. The modulator driven at 10 MHz imprints a 20 MHz RF tone on the carrier (~10 m RF-wavelength). Chromatic dispersion converts any wavelength drift of the carrier into latency variations of the received signal, i.e. creating an artifact. Tapped signals fed into a wavemeter and an O/E converter enable us to study the impact of the remaining ECL's λ-drift in measurements and serve as btb signal for the phase meter, respectively. The probe combined with a 100 GHz wide ASE gap spectrum enters the wet plant at ~ -5 dBm (Fig.1c). The ASE loading carries most of the repeater output power to ensure linear propagation of the probe, which then gets optically looped back at the other end of the cable and is received using an amplifier/ filter cascade, followed by a PIN receiver. Our commercially available phase meter samples the received probe and the btb signal at about 30 S/s. Its intrinsic noise is negligibly small. A fast polarization scrambler mitigates link PDL and PMD impacts. We assume that an optical path length change equals twice the cable length change.

## 3. SIMPLE AND ACCURATE MODELS FOR TIDES

For a qualitative understanding of tidal signatures in our data we introduce a simple earth-lunar system followed by a quantitative analysis based on a tide simulator. The moon orbits the earth, envisioned as a sphere covered with a thin water layer (Fig. 3d). Both precess around their common barycentre. Gravitational, centrifugal, and centripetal forces stand in equilibrium and deform the water surface depending on the lunar position. When they both align with the sun, it maximizes the deformation resulting in two larger water bubbles (known as spring tides). A hypothetical cable connecting the earth's North and South Pole would then experience higher than average static water pressures when passing underneath these bubbles (semidiurnal, 2x a day). At neap tide the proportions differ leading to weaker water elevations. While this simple model provides fundamental insights into the temporary evolution of the static water pressure, multiple factors, including continental boundaries, the ellipticity of the moon and sun's orbits, and the obliquity of the earth axis, all significantly impact the actual tide dynamics. NASA's GOT4.7 simulator takes the variables into account when globally computing tide levels with ~1 cm accuracy[6,7].

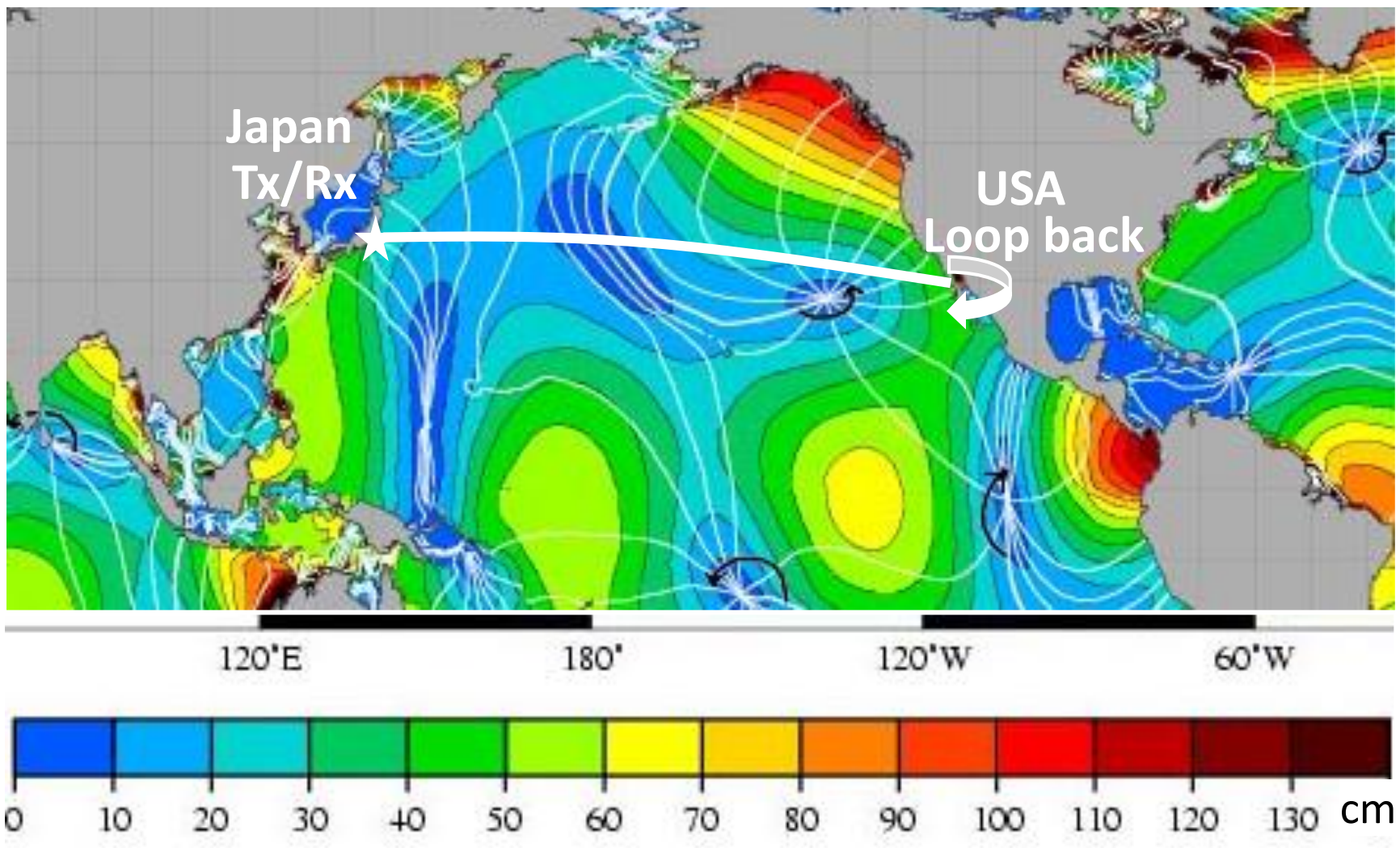

Fig.2: Dominant amplitudes (M2 constituents) for tides across the Pacific[6],[7] and approximate position of the subsea cable.

## 4. LATENCY VARIATIONS ON TRANSPACIFIC CABLE

Our continuous latency recording on a modern subsea cable connecting Japan and USA started at 2020-02-28T06:06:29 UTC, ran over ~12 days, and covered a neap tide and parts of the preceding and succeeding spring tides. The geographical path of the cable is well-approximated by an arc with minimum length on the globe's surface, which crosses areas with tidal M2 constituents of primarily low height (Fig. 2).

The probe generated in Japan was optically looped back at the other cable end in the US and the measured phase difference (MPD) to the btb signal is shown in Fig. 3a over time and relative to the aggregated tide (AT) computed with the GOT4.7.

As AT we define the integral of the local sea level elevation above normal along the cable normalized to its length. Note, at certain instants, significant local water elevations in different cable segments can cancel each other out and therefore the AT equals zero. We assume a proportionality between the local cable stretching and the corresponding water surface elevation. Hence, the overall length change linearly relates to the AT. Noise in the raw data, relating to ASE, acoustic effects, pol-scrambling, etc. were averaged over increments of 600s which suppresses high frequency components but still allows tracking of the tidal impact that appears at much lower speeds. The expected errors due to averaging are about 5%.

The MPD decomposes into a relatively strong linear tilt and a fine structure, which highly correlates with the AT. As an example, during neap tide the semidiurnal structure of both the AT and the MPD washes out, but returns at the surrounding spring tides. A linear tilt of the MPD across most of the recording does not correlate with the AT's fine structure, but rather suggests a constant shrinking of the cable. This tilt seems to flatten towards the recording's start and ending phases which coincides with new moon and full moon.

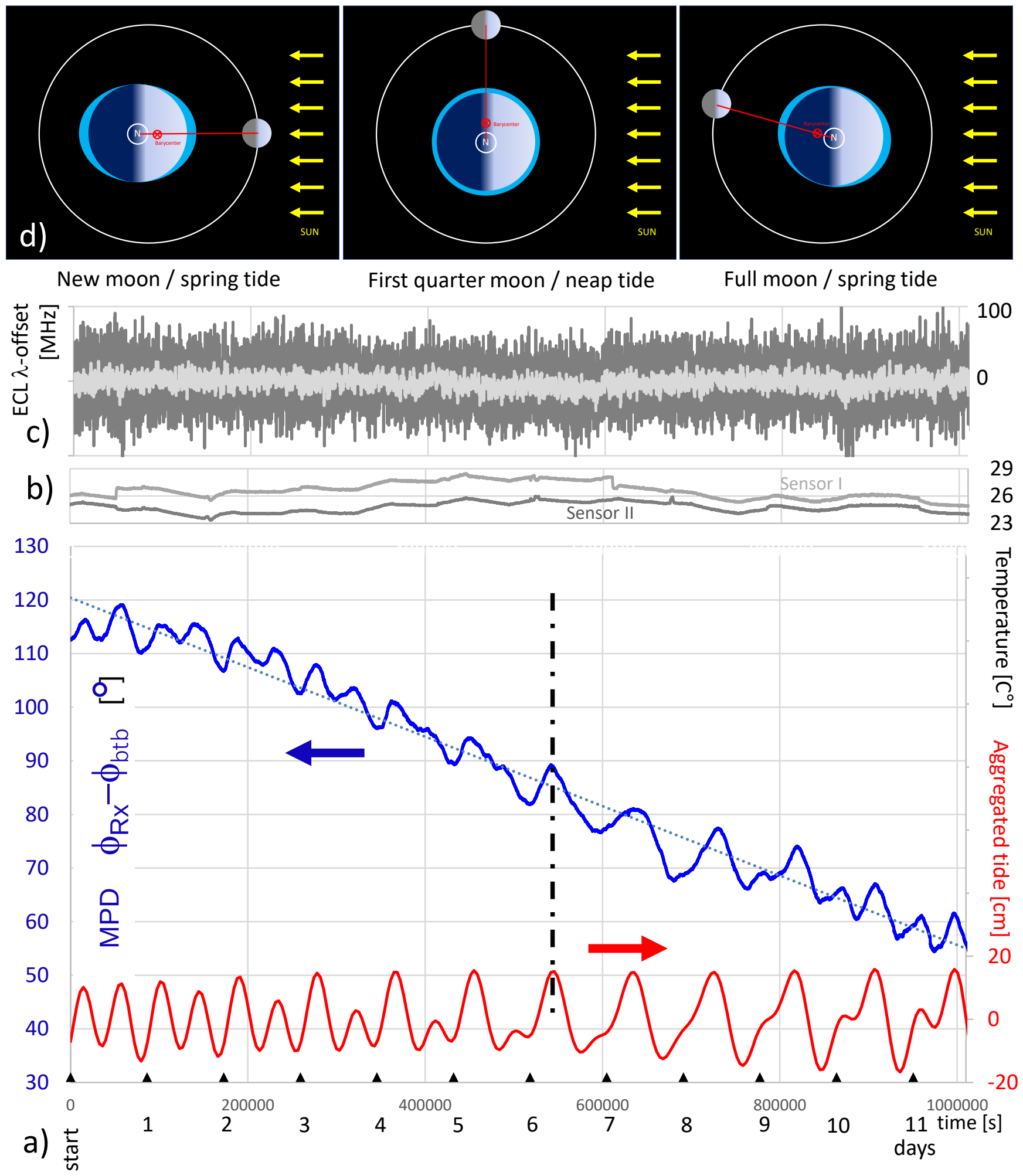

Fig. 3: a) Measured Phase Difference (MPD) over 12 days relative to aggregated tide (AT, computed with GOT4.7). b) MPD-uncorrelated and small environmental temperature changes at setup. c) Wavelength drift of ECL recorded every 60s (dark) and averaged over 600s (light). d) Simple tide model: when all objects line up maximum (spring) tides with semidiurnal period result; with the Moon perpendicular to the Sun-Earth axis, the semidiurnal tide signature weakens.

A variety of phenomena, like seasonal temperature drift along the cable, are currently being examined to explain its very small shrinking at a slew rate ~8 $10^{-14}$/s. The ECL's small λ-drift (Fig. 3c) logged every 60s ($\sigma_{ECL}$~26 MHz) and averaged over 600s ($\sigma_{mECL}$ ~8 MHz) can be clearly ruled out as the source of the medium and long-term MPD evolution, considering the fiber CD ~21 ps/km/nm. Also, short and long-term frequency instabilities of the GPS-coupled synthesizer during the signal's round-trip time and during the whole recording, respectively, contribute a negligible phase error. We estimate a variance for the phase error gene-rated on the Tx side in a range of $\sigma_{Tx}$ ~ 0.01°. No correlation has been observed between the MPD and temperature variations produced by the station's A/C units and captured with two sensors next to our setup (Fig. 3b).

## 5. POISSON EFFECT ON PRESSURIZED CABLES

The Poisson ratio from material science describes a longitudinal expansion of a body caused by lateral compression of its volume. We attribute the fine structure in the MPD to this effect. In a simplified picture, an increased static water pressure on the ocean floor due to high tide compresses the cable in a radial direction, which extends it along its longitudinal axis (Fig. 4a). Friction conveyed by gel inside the cable tube transfers the expansion of the cable jacket into fiber stretching. As the probe's propagation velocity stays nearly unaffected by fiber stretching, a length increase relates directly to a larger MPD[8]. The cross-section of modern subsea cables is mainly built of steel wires (stability), a copper conductor, and a HDPE insulation (Fig. 4b). We estimate limits for the Young modulus 'E' and Poisson ratio 'ν' of this composite by assuming the body either consists only of the steel wires or the HDPE tube. Under a pressure enhancement ΔP the length change Δl for a cylinder with an effective inner and outer radius ($r_i$, $r_o$) and a length $L_0$ reads (Lamé's Equation[9])

$$\Delta l = 2\, \nu\, E^{-1}\, r_0^2\, (r_0^2 - r_i^2)^{-1}\, L_0\, \Delta P. (1)$$

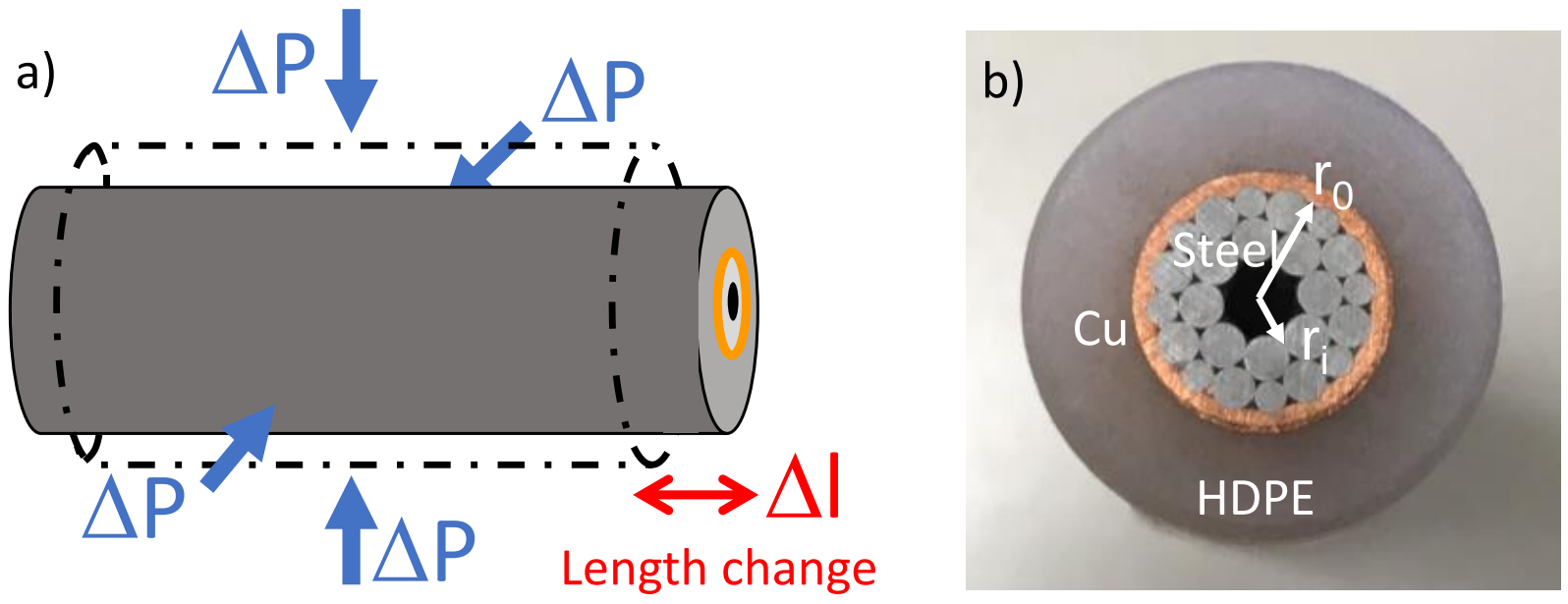

| c) | Young modulus [GPa] | Poisson ratio | $r_o$ [mm] | $r_i$ [mm] | ΔP [Pa] | 2 x Cable length [Mm] | Δl [cm] @ 8.5 cm agg. tide |
|---|---|---|---|---|---|---|---|
| Steel | 200 | 0.3 | ~4 | ~ 2.6 | 830 | ~ 20.8 | 4.5 |
| HDPE | 0.8 | 0.45 | 8.5 | 4.6 | | | 1375 |

Fig. 4: a) Poisson effect stretches cable due to enhanced lateral water pressure. b) Cross section of modern subsea cable. c) Material parameters used for Eq. 1 to estimate pressure-induced length change of pacific cable.

An avg AT level of ~8.5 cm leads to a predicted cable length change of ~4.5 cm or a MPD of 3.2° when modelling the cable as steel tube. The avg MPD after subtracting its linear phase tilt agrees surprisingly well with these estimates and reads ~3.4°. Note: for simplicity we assume the cable as a straight line with open ends for which Eq.1 holds. More accurate approximations consider the small meander shape of deployed cables and flexible, rather than rigid, coupling between cable jacket and fiber. The relative length variations a cable experiences due to tides are orders of magnitude smaller than its stretching during installation.

We performed shorter recordings on a transoceanic cable system with approx. 13.5 Mm length from May 18 to May 25, 2018 and found qualitatively similar results with respect to MPD, MPD tilt, and tide dynamics.

## 6. CONCLUSIONS

For the first time we experimentally studied arrival phase variations of an RF signal on an optically looped-back carrier between Japan and the US. Our measurement suggests that cable length variations induced by water pressure, which depends on the tides across the pacific, stretch the fiber. This contradicts a 'loose tube model' for cables. The inherent coupling between the fibers and the loose tube, though this friction is very small, allows for transmittal of tension and tinily stretches them by a few ppb. We found a strong correlation between averaged tide levels and the arrival phase over daily periods. Larger offsets appear during week-long recordings. It is surprising how long-term stable cables are but, then again, it's remarkable how much tides cycle their length on a daily basis.

**Acknowledgements**

The author would like to thank R. Ray, W. Patterson, S. Bernstein, S. Abbott, B. Bakhshi, and S. Hunziker for supporting this work.